\documentclass[aps,prd,twocolumn,groupedaddress,showpacs]{revtex4}

\usepackage{epsfig} 
\usepackage{graphics}
\usepackage{amsfonts}

\newcommand{\RR}{{\mathbb{R}}}
\newcommand{\ZZ}{{\mathbb{Z}}}
\newcommand{\pa}{\partial}
\newcommand{\tr}{\mathop{\rm tr}\nolimits}
\newcommand{\cE}{{\cal E}}
\newcommand{\cB}{{\cal B}}
\newcommand{\cA}{{\cal A}}
\newcommand{\ii}{{\rm i}}

\begin{document}

\title{A Monopole Wall}

\author{R. S. Ward}
\email[]{richard.ward@durham.ac.uk}
\affiliation{Department of Mathematical Sciences, University of
Durham, Durham DH1 3LE}

\date{\today}

\begin{abstract}
We construct, numerically, a solution of the SU(2) Bogomolny equations
corresponding to a sheet of BPS monopoles. It represents a domain wall
between a vacuum region and a region of constant energy density,
and it is the smoothed-out version of the planar sheet of Dirac monopoles
obtained by linear superposition.
\end{abstract}


\maketitle

\section{Introduction}

Bogomolny-Prasad-Sommerfield (BPS) monopoles have long been of
considerable interest (for reviews, see \cite{MS04, WY06}),
and recently have found a new interpretation in the context of D-branes.
This D-brane connection is partly responsible for a particular
interest in periodic assemblages of monopoles. For example, there have
been studies of monopole chains \cite{CK01}, where the underlying
theory (such as the Nahm transform) is
well-developed.  However, for monopole sheets or walls, where the
fields are doubly-periodic, much less is known.

It has been suggested \cite{L99, W05} that the (essentially Abelian)
homogeneous solution on $\RR^3$ may be viewed as a monopole sheet,
and there is
a sense in which such an interpretation is meaningful. If, however,
one constructs a sheet of Dirac monopoles, as a double series, then
the resulting field looks rather different. The purpose of this note
is to present numerical evidence for the existence of an SU(2)
monopole sheet which is a smoothed-out version of the
Dirac sheet. As we shall see, it resembles a domain wall between
a vacuum region and a region of homogeneous phase with constant
energy density.


\section{The U(1) Case: Sheet of Dirac Monopoles}

Let ${\bf r}=(x,y,z)$ denote the usual position vector in $\RR^3$, and
let $P$ be the square lattice in the $xy$-plane consisting of points
of the form ${\bf r}=(j,k,0)$, with $j,k\in\ZZ$. Suppose we place
a unit-charge Dirac monopole at each site of this lattice. The resulting
magnetic field is, at least formally,
\begin{equation} \label{DiracField}
 {\bf B}(x,y,z) = \frac{1}{2}\sum_{j,k\in\ZZ}(x-j,y-k,z)/r_{jk}^3,
\end{equation}
where $r_{jk}^2=(x-j)^2+(y-k)^2+z^2$. The component $B_z$ is a positive
series which converges absolutely, and the other two components $B_x$
and $B_y$ are conditionally convergent (except, of course, at the points
of $P$, where ${\bf B}$ is singular). For fixed $(x,y)$, we have
${\bf B}\to(0,0,\pm\pi)$ as $z\to\pm\infty$; this was checked by numerical
summation of the series, and can also be established by a rough analytic
approximation.

Similarly, we may obtain a gauge potential ${\bf A}$ by taking a
gauge potential for each monopole and summing these; for example,
\begin{equation} \label{DiracPotential}
 {\bf A}(x,y,z)=\sum_{j,k\in\ZZ}\frac{(k-y,x-j,0)}{2r_{jk}(z+r_{jk})},
\end{equation}
which is valid in the region $z>0$.
The asymptotic behaviour of (\ref{DiracPotential}) is
${\bf A}\to\frac{\pi}{2}(-y,x,0)$ as $z\to\infty$.


In general, if we have a doubly-periodic vector field ${\bf B}$ which is
smooth for $|z|\geq c$, then we can define its charge $(N_-,N_+)$ by
\begin{equation} \label{DiracCharge}
 N_{\pm}=\frac{1}{2\pi} \int_0^1dx \, \int_0^1dy \, B_z(x,y,\pm c);
\end{equation}
by continuity, $N_{\pm}$ do not change if $c$ is increased.
If ${\bf B}$ is genuinely a doubly-periodic magnetic field, in other words
if its gauge potential ${\bf }$ is doubly-periodic up to a gauge
transformation, then $N_-$ and $N_+$ will both be integers: this is the
Dirac quantization condition for a magnetic field on the torus $T^2$.
The field (\ref{DiracField}), in view of its asymptotic behaviour,
clearly has charge $(-\frac{1}{2},\frac{1}{2})$, and so it does not quite
satisfy this quantization condition. (Of course, taking an array of monopoles
of charge 2 would have the effect of multiplying all the expressions by 2, and
so this would give a genuinely doubly-periodic magnetic field.)

Another example of a `half-doubly-periodic' field is the homogeneous solution
\begin{equation} \label{DiracHomog}
 {\bf B}(x,y,z)=(0,0,\pi), \quad {\bf A}(x,y,z)=\frac{\pi}{2}(-y,x,0),
\end{equation}
which has charge is $(\frac{1}{2},\frac{1}{2})$. If we add
(\ref{DiracHomog})
to (\ref{DiracField}), then we obtain a doubly-periodic magnetic field of
charge $(0,1)$, with
\begin{equation} \label{DiracAsymp}
 {\bf B}\to
  \left\{
    \begin{array}{ll}
       2\pi & \mbox{as $z\to\infty$}, \\
       0    & \mbox{as $z\to-\infty$}.
    \end{array}
  \right.
\end{equation}
This is the prototype of our monopole sheet: it separates a vacuum region with
energy density approximately zero (for $z\ll-1$), from a region of homogeneous
phase with energy density approximately constant (for $z\gg1$). In what follows,
we will show that it has a non-Abelian version in which the singularities
are smoothed out.

\section{The SU(2) Case: Setup}

Let $\{A_j(x,y,z)$, $\Phi(x,y,z)\}$ denote an SU(2) Yang-Mills-Higgs field, and
define $D_j \Phi:=\pa_j \Phi +[A_j, \Phi]$,
$B_j := \frac{1}{2}\varepsilon_{jkl}(\pa_k A_l - \pa_l A_k +[A_k, A_l])$,
as usual. We impose the global conditions:
\begin{enumerate}
  \item[(a)] $A_j$ and $\Phi$ are $2\times2$ anti-Hermitian trace-free matrices,
        and are smooth on $\RR^3$;
  \item[(b)] $A_j$ and $\Phi$ are periodic in both $x$ and $y$ (actually periodic,
    not merely up to a gauge transformation), with unit periods.
\end{enumerate}
In addition, we need boundary conditions as $z\to\pm\infty$, and these can be
formulated as follows.
If the restriction $\Phi_c := \left. \Phi\right|_{z=c}$
is nowhere-zero (as a function of $x$ and $y$), then the normalized Higgs field
$\widehat{\Phi}_c := \Phi_c/|\Phi_c|$ is well-defined, and is a map from the
torus $T^2$ to the 2-sphere $S^2$. Here $|\Phi|^2 := -\frac{1}{2}\tr{(\Phi^2)}$.
So $\widehat{\Phi}_c$ has a degree (winding number)
$N_c\in\ZZ$. If $\Phi_z$ is nowhere-zero for $z\geq c$, then by continuity the
degree $N_z$ is independent of $z$, and is denoted $N_+$; similarly for $N_-$.
The boundary condition, motivated by the U(1) case, is:
\begin{enumerate}
  \item[(c)] $D_x\Phi\to0$ and $D_y\Phi\to0$ as $z\to\infty$;
    if $N_+\neq0$, then $|\Phi|/z \to 2\pi \, |N_+|$,
    and $|D_z\Phi|$ is bounded, as $z\to\infty$; if $N_+=0$,
    then $|\Phi|\to{\rm const}$ and $|D_z\Phi|\to0$ as $z\to\infty$;
    and similarly as $z\to-\infty$.
\end{enumerate}
We say that such a field has charge $(-N_-, N_+)$.

For fields satisfying conditions (a)--(c), there is a topological lower bound
on the energy. The derivation of this is analogous to the one used for monopoles
localized in $\RR^3$. In order to get finite energy, we need to restrict
to a finite cylinder $-L\leq z\leq L$; the condition (c) is adapted in the
obvious way to become a condition at $z=\pm L$.  The energy density is
\begin{equation} \label{Enden}
  {\cal E} := -\frac{1}{2}\tr\left[(D_j\Phi)^2+(B_j)^2 \right].
\end{equation}
The first observation is that
\begin{equation}
  N_c = \frac{1}{4\pi}\int_{z=c}\tr\left[\widehat{\Phi} B_z\right]\,dx\,dy,
\end{equation}
which is a standard calculation \cite{JT80}.
The energy is
\begin{eqnarray*}
 E_L &:=& \int_{-L}^Ldz\,\int dx\,dy\,\cE \\
   &=& -\frac{1}{2}\int\tr(D_j\Phi+B_j)^2\,d^3x + \int\tr(D_j\Phi\cdot B_j)\,d^3x.
\end{eqnarray*}
Assuming for simplicity that $N_+\geq0\geq N_-$, and using Stokes's theorem,
this leads to the inequality
\begin{equation} \label{BogBound}
  E_L \geq 8\pi^2 L(N_+^2 + N_-^2),
\end{equation}
with equality if and only if the Bogomolny equations
\begin{equation} \label{BogEqn}
       D_j \Phi = -B_j
\end{equation}
are satisfied.

There is an exact solution of (\ref{BogEqn}) representing a field of
charge $(1,1)$ --- in other words, with $N_+ = 1 = -N_-$.
This is obtained \cite{L99, W05} by
embedding the homogeneous Abelian field (\ref{DiracHomog}), multiplied
by a factor of -2, into SU(2):
\begin{equation} \label{NonabHomog}
  \Phi=2\ii\pi z\sigma^3, \quad A_j=\ii\pi(y,-x,0)\sigma^3;
\end{equation}
and then applying an SU(2) gauge transformation (which it is not hard to write
down explicitly), in order to make the fields periodic.
The energy density of (\ref{NonabHomog}) is easily read off, and is
$\cE=8\pi^2$; so the Bogomolny bound (\ref{BogBound}) is saturated, as expected.


\section{The SU(2) Case: Numerical Solution}

The problem is to investigate whether there is a SU(2) solution of the
Bogomolny equations (\ref{BogEqn}) with charge $(0,1)$ --- and to see
what it looks like. No existence theorem is currently available, and the
best that can
be done here is to use a numerical approach. The idea is to discretize the
system as a lattice gauge theory, and to minimize the energy $E_L$ numerically.
A minimum which saturates the lower bound (\ref{BogBound}), in other words with
$E_L=8\pi^2L$, should then be a solution.

So we replace $\RR^3$, or rather the region $(0,1]\times(0,1]\times[-L,L]$,
with a cubic $M\times M\times(2LM+1)$ lattice $\Gamma$, so the lattice spacing
is $h=1/M$. The gauge potential is represented, in the standard way, by
assigning an element $\cA$ of SU(2) to each link of $\Gamma$; the curvature
is then an element $\cB$ of SU(2) associated with each face.  We represent the
Higgs field by assigning an element $\phi$ of SU(2) to each site of $\Gamma$;
the covariant derivatives $D\phi$ are then elements of SU(2) associated with
each link. The energy is given by the `Wilson action'
\begin{equation} \label{Wilson}
  E_L = \frac{2}{h} \sum_{{\rm faces}} (1-\tr\cB)
     + \frac{2}{h} \sum_{{\rm links}} (1-\tr D\phi),
\end{equation}
which is gauge-invariant on the lattice.

The boundary conditions, which are also gauge-invariant, are as follows:
\begin{itemize}
   \item at $z=L$, we impose $\frac{1}{2}\tr\phi = \cos(2\pi hL)$;
   \item at $z=-L$, we impose $\cB=1$.
\end{itemize}
These correspond, respectively, to the conditions $|\Phi|=2\pi L$ at $z=L$,
and ${\bf B}={\bf0}$ at $z=-L$, in the continuum version.

The initial configuration was constructed by starting with the lattice
version of the homogeneous solution, namely
\begin{eqnarray*}
 \phi &=& \exp(2\pi\ii hz\sigma^3) \\
 \cA_x &=& \exp(\pi\ii hy\sigma^3) \\
 \cA_y &=& \exp(-\pi\ii hx\sigma^3) \\
 \cA_z &=& 1;
\end{eqnarray*}
then gauge-transforming so as to make these periodic; and finally adjusting
by setting $\phi=1$ for $z<0$, and interpolating so that $\cA_x$ and $\cA_y$ 
go to $1$ as $z$ goes from $0$ to $-L$.  The resulting lattice field has
charge $(0,1)$.

Starting with this initial configuration,
the energy (\ref{Wilson}) was minimized using a conjugate-gradient method,
while maintaining the boundary conditions (and also, of course, the condition
that $\cA$ and $\phi$ should be SU(2)-valued). 
This was done for various values of $M$ (or equivalently $h$) and $L$.
For $M\geq12$, the results are already of high accuracy, in the sense that
the value of $E_L$ at its minimum is within $0.5\%$ of the Bogomolny
bound, and it changes by less than that amount if $M$ is increased.
\begin{figure}[htb]
\begin{center}
\includegraphics[scale=0.6]{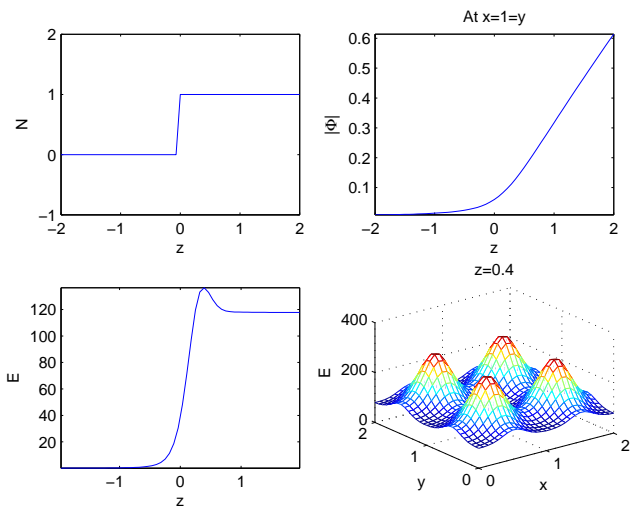}  
\caption{Charge, Higgs field and energy density of SU(2)
             monopole sheet \label{fig1}}
\end{center}
\end{figure}
The graphs in Figure 1 depict the minimum-energy configuration with
$M=14$ (so the lattice spacing is $h=0.071$), and $L=2$. Its lattice
energy is $E_L=0.9997\times8\pi^2L$.
The upper-left-hand graph shows the topological charge
(winding number) $N$ of $\left.\Phi\right|_z$, as a function of $z$;
in particular, we see that the solution does indeed have charge $(0,1)$.
The norm of the Higgs field $\Phi$ (at $x=y=1$, and as a function of $z$)
is shown in the upper-right-hand graph;
as expected, it approaches a constant value as $z\to-\infty$, as depends
linearly on $z$ as $z\to\infty$.
The lower-left-hand graph shows the energy density $\cE$ (or rather
its lattice version) summed over $x$ and $y$, as a
function of $z$.  We see that the energy density tends to zero as
$z\to-\infty$, approaches a constant value as $z\to\infty$, and is
peaked at $z\approx0.4$. Finally, the lower-right-hand subfigure plots
the energy density at $z=0.4$, as a function of $x$ and $y$. For $|z|\geq1$,
the corresponding energy plots are essentially constant in $x$ and $y$,
but at the location of the monopole sheet (which here is at $z\approx0.4$)
we see the individual monopoles
(one in each fundamental cell of the lattice, of which four are shown).

\section{Concluding Remarks and Open Questions}

It is straightforward to change the $x,y$-periods
by scaling: in fact, if $\{\Phi, A_j \}$ is a solution with unit
periods in $x$ and $y$, then
$\{\widetilde{\Phi}({\bf r}):=\lambda\Phi({\bf r}/\lambda),
\widetilde{A}_j({\bf r}):=\lambda A_j({\bf r}/\lambda) \}$
is a solution with periods $\lambda$. More generally,
it should be possible (although this has not been investigated) to construct
analogous solutions in which the monopole sheet has a different lattice
structure, for example hexagonal. (Compare the case of the Skyrme system,
where a hexagonal lattice is slightly more energetically favourable than
a square lattice \cite{BS98, MS04}.)

If one takes the gauge group to be SO(3) rather than SU(2), then the
topological classification of doubly-periodic monopoles has an additional
feature, since SO(3) bundles over the torus $T^2$ are not necessarily
trivial --- they are classified by $\ZZ_2$. Essentially, the consequence
is that the charges $(N_-,N_+)$ of SO(3) monopole sheets need not be integers,
but may be half-integers. The simplest example \cite{W05} is the
homogeneous solution of charge $(\frac{1}{2},\frac{1}{2})$, analogous
to (\ref{NonabHomog}), namely
\begin{equation} \label{NonabHomogHalf}
  \Phi^{(0)}=\ii\pi z\sigma^3,
    \quad A_j^{(0)}=\frac{1}{2}\ii\pi(y,-x,0)\sigma^3;
\end{equation}
which, in effect, lives on a non-trivial SO(3) bundle over
$T^2\times\RR$. There should exist an SO(3) domain wall solution of charge
$(0,\frac{1}{2})$ which is analogous to the charge $(0,1)$ solution.
More generally, there should exist sheets of charge $(p/2,q/2)$, where
$p,q\in\ZZ$.  If $p$ and $q$ are distinct, then such a solution
represents a wall between two distinct homogeneous phases; one of these
could be the vacuum, if the corresponding charge is zero.

The only `visible' parameters in the numerical solution presented above are
those corresponding to translations in $x$, $y$ and $z$. (The position of the
sheet, which turned out to be at $z\approx0.4$, is in effect set by the
boundary condition at $z=L$: in fact, by the fixed value of $|\Phi|$ at $z=L$.)
Whether there are additional moduli is not known. For the homogeneous
solutions, the moduli can be computed explicitly \cite{L99, W05}: the
solutions of charge $(p/2,p/2)$, with $p\in\ZZ$, depend on $4p$ parameters;
and the perturbations (normalizable zero-modes) can be written down explicitly
in terms of theta-functions. (Only the $p=1$ case was presented in \cite{W05},
but its generalization to $p\geq2$ is straightforward.) This suggests that there
might also be additional moduli in the general case of charge $(p/2,q/2)$.

The main tool in the analysis of BPS monopoles has been the Nahm transform
\cite{N82, CG84}. The general pattern \cite{J04} suggests that the Nahm transform
of a monopole sheet (a solution on $T^2\times\RR$), will be
another solution on $T^2\times\RR$ --- in other words, that monopole sheets
are `self-reciprocal' under the Nahm transform.
The only current evidence in favour of this
comes from the class of homogeneous solutions: the U(1) case was described in
\cite{W05}, and the SU(2) case is similar. 

Solutions (with prescribed singularities) of the Bogomolny equations on
$C\times I$, where $C$ is a Riemann
surface and $I$ is an interval, crop up in the area of supersymmetric
gauge theory and branes (see, for example, \cite{KW06}). The context
of the present paper is rather different, but the solutions described here
could be re-interpreted in D-brane language.

There are many similarities between BPS monoples and Skyrmions (see, for
example, \cite{HMS98}). In the Skyrme system, there is a wall-like solution
analogous to a graphene sheet, and one may view fullerene-like Skyrmions as
being shells constructed from such sheets \cite{BS98}; in fact, the sheet
separates a region of `Skyrmion core' from the region outside the Skyrmion.
The monopole sheet described above appears at first sight to be rather
different in nature, since neither side of the wall corresponds to the
field outside a BPS monopole (which, of course, is that of a Dirac monopole).
It may, however, turn out to be relevant as the wall of a magnetic bag
\cite{B06}.

To summarize: we have constructed, numerically, a solution of the Bogomolny
equations representing a sheet of BPS monopoles: in general, it is a domain
wall between two regions of constant energy density, either of which
can be a vacuum region.

Note added: I thank the referee for emphasizing that a purely-magnetic wall
with vacuum on one side and a nonzero magnetic field on the other,
cannot be static: it would accelerate in the direction of the
magnetic field. In the BPS case, by contrast, the pressure on
the wall from the Higgs field is exactly opposite to the pressure
from the magnetic field, and so there is no obvious instability.


\begin{acknowledgments}
This work was supported by research grants ``Strongly Coupled Phenomena''
and ``Classical Lattice Field Theory'',
from the UK Particle Physics and Astronomy Research Council.
\end{acknowledgments}


\end{document}